# SEMI-PHENOMENOLOGICAL MODEL
# FOR A WIND-DRIFT CURRENT

**Vladislav G. Polnikov**

A.M. Obukhov Institute of Atmospheric Physics of RAS, Moscow, Russia;

polnikov@mail.ru

## Abstract

Some data of the drift current, $U_d$, measured on a wavy surface of water in a laboratory and the field, are briefly described. Empirical formulas for $U_d$ are given, and their incompleteness is noted, regarding to absence of the drift current dependence on surface-wave parameters. With the purpose of theoretical justification of empirical formulas, a semi-phenomenological model of the phenomenon is constructed. It is basing on the known theoretical and empirical data about the three-layer structure of the air-water interface: 1) the air boundary layer, 2) the wave-zone, and 3) the water upper layer. It is shown that a presence of linear drift-current profile in the wave-zone, $U(z)$, makes it possible to obtain a general formula for the drift current on the wavy water surface, $U_d$. This model is based on the balance equation for the momentum-flux and current-gradient, taking place in the wave-zone. The proposed approach allows us to give an interpretation of the empirical results, and indicate the direction of their further detailed specification.

## 1. Introduction

Presence of air current (wind) above a calm surface of water leads very quickly to an appearance of waves and the drift current, $U_d$, on a surface of water. The stable wind profile, $W(z)$, is formed in the air boundary layer and the drift velocity profile, $U_d(z)$, do in the water upper layer, depending on the state of surface waves. All these phenomena are caused by the appearance of a turbulent momentum flux from the wind to the interface. This flux is often referred to as the wind stress, $\tau$. For manifestation of these phenomena, it is required to get the regime of sufficiently high Reynolds number, $Re = WL/\nu_a$, (where, $L$ is the spatial scale of the wind-variability, $\nu_a$ is the kinematic viscosity of air), i.e. the wind speed should excess a certain threshold value. According to experiments in the tanks, it is 2-3 m /s (Monin & Yaglom, 1971; Phillips, 1977; Wu, 1975; Longo et al., 2012a).

Due to the simultaneous combination of shear flows, wavy and turbulent motions, taking place in the vicinity of moving air-water boundary (hereinafter referred to as the interface), the hydrodynamics of the interface system is very complicated (Monin & Yaglom 1977; Phillips 1978). In Section 3, it will be shown that the entire system of interface can be rather clearly shared into three constituent parts: the air boundary layer (ABL), where the air is permanently present; the wave-zone (WZ), where the air and water are alternately present; and the upper water layer (WUL), where water is always present[1]. In the description dynamics of the interface system, information about the wind-wave state is very significant, due to the fact that wind-waves are the mediator of all the movements near interface. However, here we will consider the parameters of ABL and WZ as given ones, and confine ourselves to the task of constructing a theoretical description of wind-drift current $U_d$ over the WUL, as a function of the wind speed, $W$, and the wave parameters nominated below. This is the purpose of this work.

At present time, there is no theoretical model that allows giving the description of wind-drift current, mentioned above. The solution of this problem assumes extensive using all the available experimental data about the relation of drift-current $U_d$ with both the wind and waves parameters and the structure of interface. This kind of information on the structure of interface was established only in recent years (see below), what opens up an opportunity for solving the task.

Despite the fact that values of $U_d$ are small in comparison with the wind speed at the standard horizon $z$, $W_z$, or with the celerity of wind waves, the theoretical description of wind-drift is of both physical and practical interest. The first of them is determined by the necessity to clarify the nature of phenomenon, and the latter is important for solving the problems of navigation and marine activities safety, and for managing environmental problems (estimating the rate and region of impurity distribution on the water surface) (Wu, 1975; Longo et al., 2012a; Churchil & Csanady, 1983; Babanin, 1988; Malinovskii et al, 2007; Kudryavtsev et al., 2008; among others).

---

[1] We mean that the layer is a horizontally extended region of space, for which at least one of the vertical boundaries is not well defined (ABL, WUL), whilst the zone is a layer, both vertical boundaries of which are defined.

Experimental studies of the wind-drift are very numerous (see, for example, references in the papers mentioned above), what is explained by availability of the object of research and simplicity of measurement for it. However, this simplicity, in fact, is only apparent, since accurate measurements on wavy (oscillating) interface are far from to be easy. Indeed, when measuring drift velocities in the field (for example, Churchil & Csanady, 1983; Babanin, 1988; Malinovskii et al, 2007; Kudryavtsev et al., 2008), there are appearing a lot sources for errors: nonstationarity of wind speed $W$, uncontrolled background currents $U_b$ in the measurement area; uncontrolled extent of stratification for air in the ABL and water in the WUL, and so on. At the same time, the advantage of field-measurements is a very wide range of realization for the wind-wave conditions.

An alternative method for studying the regularities of drift currents formation is based on laboratory (tank) measurements. But there are drawbacks in this approach. Indeed, in the tank measurements (see the examples in Wu, 1975; Longo et al., 2012a), the causes of errors are an influence of the vertical and lateral boundaries of the tank, the presence of reverse currents, the small scale of wind- and wave- fetch, and so on. All these limitations affect on the kind of wind profile, and, consequently, on establishing the corresponding dependencies of $U_d$ on parameters of the ABL. Small dimensions of tanks are decreasing the possibility of establishing dependences $U_d$ on the wave state, in a wide range of its parameters. However, the tank measurements have an important advantage, as far as all the parameters of wind-wave conditions are completely controlled during such an experiment. In this regard, it is the ideal way to carry out experiments in tanks with transverse dimensions of the order of a meter and a length of several tens of meters (Huang & Long, 1980).

The listed and other sources of the errors impose significant limitations on the accuracy of results measured. Herewith, the experiments themselves require a careful preparation both for the measurements and their analysis, which is described in detail in the cited references. Although estimates of the accuracy of measuring drift velocity $U_d$ are not always presented, an analysis of scattering the published data allows us to assume that the measurement errors for $U_d$ are about 10% (see, for example, figures in Wu, 1975; Longo et al., 2012a; Churchil & Csanady, 1983; Babanin, 1988; Malinovskii et al, 2007; Kudryavtsev et al., 2008). This errors value is also confirmed by direct the estimates of measurement errors in (Babanin, 1988; Malinovskii et al, 2007).

Before representing the empirical formulas for drift velocity $U_d$ and information on the interface structure, needed for constructing the model, we note that in the absence of background current $U_b$, the velocity vector of the surface current on a wavy surface, $\mathbf{U}_S$, includes, in fact, two terms:

$$\mathbf{U}_S = \alpha_1 \mathbf{U}_{St} + \alpha_2 \mathbf{U}_d. \qquad (1)$$

Here $\alpha_1$ and $\alpha_2$ are the proportionality coefficients close to unity (Malinovskii et al, 2007), and $\mathbf{U}_{St}$ is the Stokes-drift vector, which is different for each of spectral components of surface waves and directed along the propagation-vector for each of these components (Stokes, 1847). As is known, the Stokes drift (transport of the liquid particles along the direction of wave propagation) is created by the uncloseness of the orbits for nonlinear wave motions, i.e. it is due only to the wave processes and is not directly related to the wind. Therefore, the Stokes drift is additive one to the wind-drift, $\mathbf{U}_d$, whose vector coincides with the wind-stress vector, $\boldsymbol{\tau}$.

Note, however, that, in the field experiments, the direction of wind-stress vector $\boldsymbol{\tau}$, and the vector of wind-drift, $\mathbf{U}_d$, may not coincide with the direction of local wind $\mathbf{W}$ (Babanin, 1988; Malinovskii et al, 2007; Kudryavtsev et al., 2008). This effect is due to the influence of Coriolis forces (realized on the spatial scales of hundreds kilometers), associated with the rotation of the Earth. Further in this paper, this effect will not be considered, what means adopting the approximation of the "spatial locality" for the air-sea interaction processes on the wavy water surface.

The expression for the magnitude of current $U_{St}$, caused by a nonlinear gravitational wave with amplitude $a$, frequency $\omega$, and wave number $k$, was obtained by Stokes still in 1847, and it is well known (Phillips, 1977; Stokes, 1847):

$$U_{St} = (\omega / k)(ka)^2 = (\omega a)(ka). \qquad (2)$$

Here two forms for representing $U_{St}$ are given, showing the relationship between the Stokes drift, the slope of wave, $\varepsilon = ka$, and the horizontal phase velocity of the gravitational wave, $c_{ph} = \omega / k$ (the first equality), and with the modulus for vertical velocity of fluid particles, $u_3 = \omega a$ (the second equality). As shown in (Wu, 1975; Longo et al., 2012a; Churchil & Csanady, 1983), the value of $U_{St}$ has approximately 10-15% of wind-drift $U_d$, what indicates the necessity to take it into account in measurements and practical problems, especially in the presence of intense long waves having a high phase velocity (for example, in storms). However, further, in the view of additivity for the terms in (1), and in the context of the problems that we solve, we do not need to take into account the Stokes drift in this paper.

## 2. Empirical data and analysis

In our problem, there are important experimental data concerning both the empirical formulas for the drift velocity on water surface, $U_d$, and the measured features of mass-transfer velocity profiles in all three parts of the interface: in the air-layer, in the wave-zone, and in the water-layer. Leaving description of the interface structure and the wind- and current-profiles for the subsequent subsection, first, we discuss the empirical formulas for wind-drift velocity $U_d$. Herewith, we note that all the results under consideration are referring to the long-term averaged drift-velocity on the water surface, the level of which is not specified.

### 2a. Formulas for $U_d$

Let's start from the measurements of wind-drift velocity, performed in laboratory tanks under strictly controlled conditions (see references in Wu, 1975; Longo et al., 2012a). In the well-known work by Wu (1975), which became the classical one, a simple linear relation was found

$$U_d = \alpha_d \cdot u_{*a}, \tag{3}$$

where $u_{*a}$ is the friction velocity in the ABL, and $\alpha_d \approx 0.53$. The value of $u_{*a}$ was established in Wu (1975) with a standard method, by measuring the wind velocity values $W(z)$ at a number of horizons $z$ located far from the mean interface-line, on the basis of the well-known formula for the logarithmic law (Monin & Yaglom, 1971; Phillips, 1977):

$$W(z) = (u_{*a}/\kappa)\ln(z/z_0). \tag{4}$$

It is important to note that in paper Wu (1975), there was not established a direct dependence of $U_d$ on wave parameters: for example, the average wave-height, $H \approx 2<a>$ ($<a>$ is the variance of free surface elevations $\eta(t)$), the average wave slope, $\varepsilon = k_p<a>$, and the age of waves, $A$, defined by the relation

$$A = c_{ph}(\omega_p)/W_{10}. \tag{5}$$

In (5), $c_{ph}(\omega)$ is the phase velocity as function of frequency, taken at the peak frequency of wind-wave spectrum, $\omega_p$; and $W_{10}$ is the wind speed at the standard horizon, z=10 m. Herewith, in Wu (1975) it was clearly shown that the drift velocity, $U_d$, decreases with the wave-fetch increasing, what is in a good agreement with the known variability of the friction velocity (see survey of data in (Polnikov et al., 2003) and detailed parametrization in (Polnikov, 2013)).

Formulas for $U_d$, similar to (3), were obtained in all other experimental works both in tanks (Longo et al., 2012a, and references therein), and in field conditions (Babanin, 1988;

Malinovskii et al, 2007; Kudryavtsev et al., 2008). These formulas differ only by coefficient $\alpha_d$ on the right-hand side of (3). For example, in the recent paper by Longo at al. (2012a), where the most modern laboratory measuring equipment was used, the value of coefficient $\alpha_d \approx 0.4$ was established. Herewith, in this case, the strange growth of friction velocity $u_{*a}$ with the fetch was fixed, accompanied by the proper growth of $U_d$; although, as it should be, the steepness of waves fell down with the fetch (see Tables 1 and 6 in Longo (2012)). In order to explain these effects, it was noted in Longo at al. (2012a) that all the differences in the laboratory measurements of $U_d$, which obtained by different authors, are simply related to the geometry of the tanks, where the experiments are performed. In the view of the above remarks about the drawbacks of tank experiments, here one can agree with the this interpretation given by the authors of the paper mentioned, and proceed to the results of field experiments.

In the field measurements (for example, Tsahalis, 1979; Babanin, 1988; Malinovskii et al, 2007; Kudryavtsev et al., 2008), the same formula (3) was established for the drift velocity, though the values of $\alpha_d$ are varying within a wide range: from 0.24 in (Babanin, 1988) to 1.5 (Tsahalis, 1979). Such a wide scattering is due to both the natural variability of the wind-wave conditions and the technical difficulties of performing accurate measurements in the field experiments. The last reason, apparently, explains the lack of information about the direct dependences of $U_d$ on the surface-wave parameters, which could be realized under the field conditions. However, it should be noted that both a significant scattering values of coefficient $\alpha_d$ and no monotonic dependence of ratio $U_d/u_{*a}$ on $W$ (Wu,1975) can mean an existence of certain (though "hidden") dependence of $\alpha_d$ on wave parameters, which is not established experimentally yet.

It can be assumed that all the dependences of drift velocity on the wave state are "hidden" in the direct proportionality between $U_d$ and friction velocity $u_{*a}$, whilst the latter, as is well known, depends explicitly on the above-mentioned wave parameters: $H$, $\varepsilon$ and $A$ (see references in Polnikov et al. (2003), Polnikov (2013)). However, the absence of direct empirical dependences of $U_d$ on wave parameters, in our opinion, requires its justification basing on specialized experiments for their determination.

*2b. The structure of interface and wind-current-profiles*

Here, first of all, it is necessary to represent the facts confirming both the existence of a three-layer interface structure, mentioned in the introduction, and the physical expediency of accounting such a structure in further theoretical constructions.

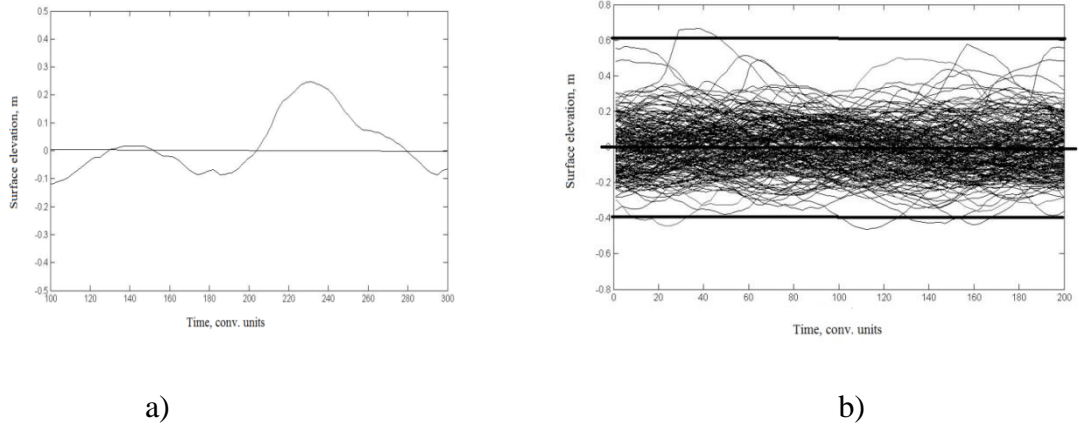

a)                                                   b)

Fig. 1. (a) The element of wave record $\eta(t)$ (the time scale is given in conventional units).
(b) The ensemble of two hundred segments of wave record $\eta(t)$ of the same realization.
The solid lines show: in (a) the conditional mean water-surface level;
in (b) the conditional boundaries of the wave-zone.

First, in order to visualize the fact of existing the wave-zone, (WZ), which should have its own physical properties, we represent here figures 1(a, b) taken from (Polnikov, 2010, 2011). In Fig. 1a, the single time realization of the free-surface elevation of water, $\eta(t)$, is shown, on the background of which any experiment is usually performed. Figure 1b shows the ensemble of two hundred such realizations of the same length, taken at the same point. Such an ensemble corresponds to the time scale of the order of hundreds of periods for the dominant wave (corresponding to the peak of wave spectrum). On this time-scale all average values of the system under consideration are described, including the wave spectrum. As can be seen from Fig. 1b, on such a scale, in the vicinity of the mean water level, the continuous zone is formed (WZ), in which the air and water are alternately present. It is natural to assume that this zone is an independent element of the interface system, the physical characteristics of which, on the mentioned time scales, should be described in a unified manner.

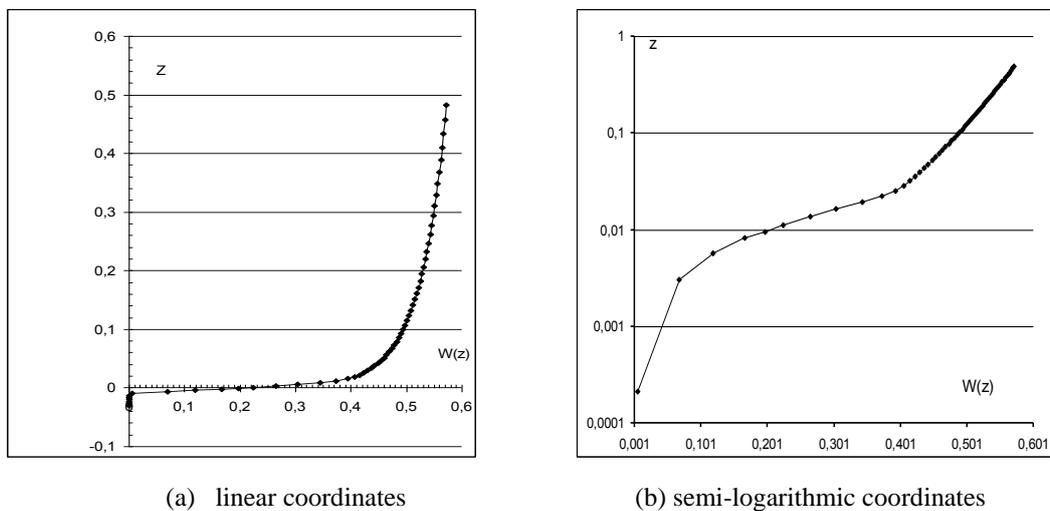

(a) linear coordinates                          (b) semi-logarithmic coordinates
Fig. 2. Calculated profile of the mean wind $W(z)$ over a wavy water surface (from Polnikov, 2011). The results are given in dimensionless quantities, and zero value of $z$ corresponds to the mean water level.

Second, as it was demonstrated for the first time in Polnikov(2011), basing on the analysis of numerical simulations executed in (Chalikov & Rinechik, 2011), the profile of average wind-velocity, *W(z)*, differs in the wave-zone significantly from the traditional logarithmic one. It varies linearly with the height from level $z \approx -h$ to level *z* of the order of $3h$, relative to the mean water level (Fig. 2) (where $h = <a>$ is the variance of water-surface deviations).

Third, these calculations are fully supported by the Longo's measurements (Longo et al, 2012b) (Fig. 3). It can be clearly seen from Fig. 3 that, in a wide range of horizons above the mean water level, and in a somewhat smaller region below, the linear profiles are really observed, both for wind speed *W(z)* and for current velocity *U(z)* (in our notations). Outside this region, which, in fact, determines the WZ, the wind profiles, *W(z)* in the ABL, and current *U(z)* in the WUL, get the form close to the wall-turbulence logarithmic profiles of form (4), having their own parameters $u_*$ and $z_0$ in each of them (see the details in Longo et al, 2012b).

Regarding the physical expediency of introducing the WZ, as the separate element of interface, it is due to the fact that dependencies of the mean wind and current on *z*, *W(z)* and *U(z)*, have the linear profiles. From the hydrodynamics point of view, such velocity profiles are the genuine characteristic for regions of viscous flows supported by viscosity coefficient *K* independent of *z*. Due to the turbulent nature of motions in WZ, this quantity, *K*, should be considered as the turbulent viscosity, the effective formula for which is the object of theoretical constructions.

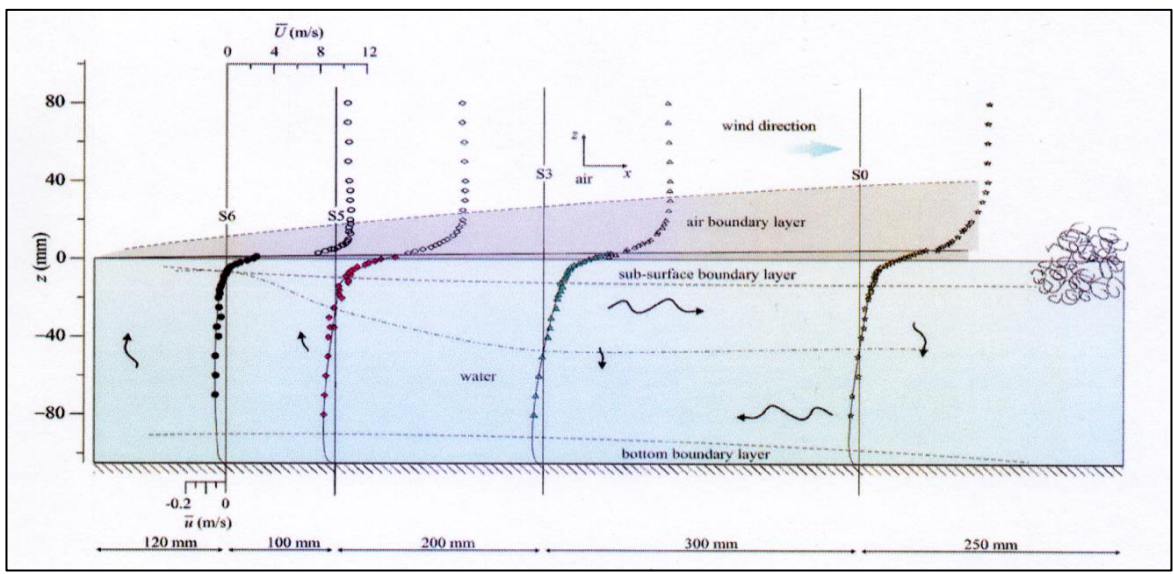

Fig. 3. General scheme for the mean flows distribution in the interface system (from Longo et al, 2012b). The speed scale for the ABL is shown at the top ($\bar{U}$, m/s), for the WUL does at the bottom ($\bar{u}$, m/s).

Thus, summing up the facts presented in this subsection, we can state the following conclusions: 1) the wavy interface system has the three-layer structure, including: ABL, WZ, and WUL; 2) from the hydrodynamic point of view, the WZ is an analog of friction layer, having linear profiles for mean velocity (both wind and current), located between the ABL and WUL; 3) in ABL and WUL, their own wall-turbulence profiles of form (4) are realized for mean wind and current, with parameters determined by the presence of wind waves on the free surface.

*3. The model for a wind-drift current*

*3a. Basic grounds*

On the basis of empirical data mentioned in the previous section, a physical model, allowing justifying the observed relation of form (2), can be constructed be the following mean.

Let us start from the fact that in the ABL there is the wind-induced downward momentum-flux (or wind stress) $\tau_a$, going from the wind in the ABL to WZ, and further into the WUL. This flux could be written in the form

$$\tau_a = <w_1 w_3> = u_{*a}^2, \tag{6}$$

where, on the right-hand side, there is the friction velocity in the air, $u_{*a}$, to the second power (formula (6) is, actually, the definition of $u_{*a}$). Here and further, the wind stress, $\tau_a$, is written terms of its modulus and in normalization to the air density $\rho_a$; whilst the wind velocity components, $w_1$, $w_3$, correspond to wind-vector representation as $\mathbf{W} = (w_1, w_2, w_3)$. Angular brackets, as usual, mean the averaging over a statistical ensemble.

It is physically clear (for the justification, see, for example, Janssen, 1991; Polnikov, 2011) that only the certain part, $\tau_t$, (called as the "skin drag") of the total flux, $\tau_a$, is spent on creating a surface current, i.e. the wind drift. The other part, $\tau_w$, ("form drag") of the total flux is expended on the growth of wave energy. Thus, one can write

$$\tau_a = \tau_t + \tau_w. \tag{7}$$

In works of the author (Polnikov, 2011, 2013), where quantity $\tau_t$ is called as "the tangential component" of the total flux, the model was proposed for calculating magnitudes of each terms in (7), as functions of wind speed $W_{10}$ and previously mentioned wave parameters. This model is not used here, but it is important to note that value $\tau_t$ can be further considered as the known one. According to estimates of paper (Polnikov, 2013), it ranges from 40 to 60% of total flux $\tau_a$, and this proportion depends on the wave state.

Thus, flux $\tau_a$ is coming from the ABL to the WZ, where it is shared into two mentioned parts. In turn, a part of the energy acquired by the waves in the WZ is carried by waves away, due to their progressive feature, and some part of the wave energy dissipates immediately with the rate $E_{wd}$, transmitting the momentum fluxes to both the WZ and the WUL. In our understanding (Polnikov, 2013), the main mechanism for the wind-wave dissipation is precisely the interaction of waves with the turbulence in WZ and WUL. Herewith, the nature of this turbulence is not important, since it is clear that the turbulence is generated by all possible mechanisms: breaking of crests; shear instability of mean currents and orbital wave motions; pulsations of the air pressure; formation of droplets and bubbles in WZ and WUL, etc. (for a discussion of alternative mechanisms of wave energy dissipation, see Babanin, 2009).

A part of the dissipating energy flux, $E_{wd1}$, is lost in the WZ, and the remainder part, $E_{wd2}$, does in the whole WUL, which extends up to the depth of about a half of the dominant wave length (due to the exponential feature of decaying amplitude for wave-spectrum components: $a(k,z) \propto exp(kz)$) (Phillips, 1977). Here, we will assume that the second part of the dissipative flux, $E_{wd2}$ is completely transmitted to the turbulence of the WUL. The first part, $E_{wd1}$, by the matter of physics, generates the turbulent viscosity in the WZ, and, possibly, in some small fraction of it, transmits some momentum to the drift current (for example, due to breaking processes). Consequently, the final value of the momentum flux, transmitted to currents in the interface system, $\tau_t$, should be slightly higher than the value mentioned above. However, since today there is no clear assessment of an additional inflow of the momentum from the dissipating waves to the drift current, we assume that the tangential flux of the horizontal momentum, $\tau_t$, realized at the upper boundary of the WZ, has the following quantity

$$\tau_t \approx 0.5 \cdot \tau_a = 0.5 \cdot u_{*a}^2. \tag{8}$$

Further, we assume that it is momentum (8) does form the wind drift both in the WZ and in the WUL (though, the exact digit in (8) does not any principal role).

Note, however, that in order to determine the drift velocity in the water, including WZ, it should be used the tangential momentum flux, $\tau_{tw}$, normalized to the density of water, $\rho_w$. Therefore, by virtue of the continuity of tangential part of momentum flux through the interface, the value of flux in water, $\tau_{tw}$, is given by the ratio

$$\tau_{tw} = \rho_a \tau_t / \rho_w = ro \cdot 0.5 \cdot u_{*a}^2, \tag{9}$$

where $ro = \rho_a / \rho_w \approx 10^{-3}$ is the ratio of air and water densities. Formula (9) also determines the friction velocity in water, $u_{*w}$, by the ratios

$$u_{*w}^2 \equiv \tau_{tw} = ro \cdot \tau_t \approx ro \cdot 0.5 \cdot u_{*a}^2 . \tag{10}$$

Now, let us clarify the geometry of vertical distribution of the drift current. On the basis of the observation results, presented in Longo et al. (2012b) and shown in Fig 3, it can be stated that the drift current localized in the WZ has a linear profile. At the upper boundary of the WZ, the sought surface drift, $U_{d0}$, takes place (further, the lower index "0" will be omitted, for simplicity of designation). It is this value is considered as the observation result. And at the lower boundary of the WZ, the drift velocity should be determined by the friction velocity in water, $u_{*w}$, given by formula (10). Velocity $u_{*w}$ is also used in the formula of logarithmic profile of form (4) for the WUL, starting from the lower boundary of the WZ. According to (10), $u_{*w} \approx (ro/2)^{1/2} u_{*a} = 0.02 u_{*a}$, it is always valid the inequality: $u_{*w} << u_{*a}$, and, consequently, it is valid that $u_{*w} << U_d$, what is very important for our aims.

Then, to close the momentum flux, we use the standard method of the turbulence theory (e.g., Monin & Yaglom, 1971; Phillips, 1977; Wu, 1975; Longo et al., 2012a), according to which the constant vertical momentum flux, $\tau_{tw}$, in the layer of viscous flow (the analogue of which is the WZ, as was shown above) is balanced by the vertical gradient of mean velocity, i.e.

$$\tau_{tw} = K_t \frac{\partial U_d(z)}{\partial z} . \tag{11}$$

Here, $K_t$ is an unknown turbulent viscosity, taking place in the WZ, the value of which is provided by the whole complicated hydrodynamics of this zone. To complete the construction of model, it remains to add that: a) the value of $K_t$ is constant in the height; b) the vertical velocity gradient in (11) is also constant, according to the experimental data.

*3b. Semi-phenomenological approximation of the model*

According to the said above about profile $U(z)$ in the WZ, with a high degree of reliability, the estimate of the velocity gradient in (11) is given by

$$\frac{\partial U_d}{\partial z} = (U_{d0} - u_{*w})/(c_{zw} h) \approx U_d / 2h , \tag{12}$$

where $h = <a>$ is the average wave amplitude at the surface-point under consideration; and, in the final expression of (12), index "0" is omitted, with the aim of simplicity. In the first, exact equality, the dimensionless coefficient, $c_{zw}$, is formally introduced, the value of which follows from the measurements. In the final, approximate equality, the height of WZ is taken from the results of (Longo et al., 2012b) in the explicit form.

The turbulent viscosity function $K_t$ is easy to be parameterized, as is customary used (Monin & Yaglom, 1971; Phillips, 1977; Polnikov, 2011,2013), in the frame of dimensional considerations. Here we state that: 1) the average wave height, $h$, is the characteristic size; and 2) the value of drift current at the upper boundary of the WZ, $U_d$, is the characteristic velocity. In this case, we obtain the expression

$$K_t = f_t(h, \varepsilon, A, \ldots) \cdot U_d h, \qquad (13)$$

in which there is the unknown dimensionless function, $f_t(h, \varepsilon, A, \ldots)$, depending only on the wave parameters. As a result, the balance (11) acquires the form of an equation for determining drift velocity $U_d$ on the upper boundary of the WZ:

$$ro \cdot 0.5 u_{*a}^2(S, W_{10}) = (f_t U_d h) \cdot (U_d / 2h) = f_t(h, \varepsilon, A, \ldots) U_d^2 / 2. \qquad (14)$$

Here, in the left-hand side, the known dependence of friction velocity, $u_{*a}(S, W_{10})$, on wave spectrum $S$ and wind speed $W_{10}$ (for example, see Polnikov et al, 2003, Polnikov, 2013) is noted for completeness of description, and, in the right-hand side, a potential dependence of unknown function $f_t$ on the wave parameters is noted as $f_t(h, \varepsilon, A, \ldots)$, which will be symbolically denoted later as $f_t(.)$.

From equation (14), one can immediately found the solution for the sought drift velocity on the wavy surface of water in the form

$$U_d = [ro / f_t(.)]^{1/2} \cdot u_{*a}. \qquad (15)$$

From the above, it is clearly seen that result (15) coincides with the observations: in the range of values $U_d \approx (0.3 \div 0.5) \cdot u_{a*}$), the values of $f_t(.)$ should be varying within the range: $f_t(.) \approx$ 0.012 - 0.02, including a possible dependence of $f_t(.)$ on the wave parameters. Such an estimate of values for coefficient $f_t(.)$, introduced in formula (13), is quite plausible, since in the practice of studying boundary phenomena there is a large number of dimensionless quantities of such order. For example, the Phillips' parameter, $\alpha_{Ph}$, standing in the formula for intensity of the tail for a saturated wind-wave spectrum (Phillips, 1977):

$$S(\omega) = \alpha_{Ph} g^2 \omega^{-5}, \qquad (16)$$

has the magnitude of the order of $\alpha_{Ph} \approx 0.01$. Another examples is the Charnock's parameter, $\alpha_{Ch}$, defined by the formula

$$\alpha_{Ch} = z_0 / ( u_{*a}^2 / g ), \tag{17}$$

which has the same order (Phillips, 1977). As can be seen from (17), $\alpha_{Ch}$ is the dimensionless characteristic for the roughness height, $z_0$, used in the formula of logarithmic layer (4), written in the terms of friction velocity $u_{*a}$ and gravity acceleration $g$. Moreover, it is well known that both dimensionless values in (16), (17) are the functions of wave parameters and wave-formation factors, and their magnitudes can vary within a wide range (Phillips, 1977).

The foregoing gives the sufficient grounds for the stating that formula (15) is the theoretical justification for empirical formula of form (3), what completes, in general, the solution of the problem posed.

### 4. Discussion of the results

First of all, it is interesting to consider an alternative solving the point of parametrization for the turbulent viscosity function, $K_t$. Let us take the friction velocity in water, $u_{*w}$, as the scale for velocity (instead of $U_d$), changing correspondingly unknown function $f_t(.)$ by the other unknown one, $f_{tw}(.)$. Then, it is easy to obtain from (10) and (13) that the functional representation of $U_d$ gets the form: $U_d \sim ro^{1/2} u_{*a} / f_{tw}(.)$. By changing the notation of unknown, $f_{tw}(.)$, by other the unknown, $f_t^{1/2}(.)$, one can state that the latter representation of $U_d$ does quite correspond to ratio (15). Therefore, further discussion of the result obtained will be performed on the basis of formula (15).

It is also interesting to analyze an applicability of ratio (15) in the limiting case of complete absence of waves. In this case, there is no WZ, the vertical gradient of drift velocity and the introduction of viscosity function become to be not needed, and function $f_t(.)$ degenerates into a unity. Then, from (15) it follows that velocity $U_d$ is simply equal to the friction velocity in water, $u_{*w}$, which is determined by formula (10) without coefficient 0.5. In other words, the functional form of formula (15) is preserved, what allows us to analyze it further.

According to observations (see Section 2), the drift velocity at the surface of water, $U_d$, depends only on the friction velocity in the air. The explicit dependences of $U_d$ on the wave parameters are not established in the experiments yet (till now). This can be stipulated by either the physics of processes considered (what, as shown above, is fully treated theoretically on the basis of formula (15) under the condition of invariability of $f_t(.)$ ) or the lack of proper experimental data. In this regard, model (15) gives the possibility to make a more detailed description of the drift current properties.

In particular, let's try to answer the question: what dependences of the drift current on the parameters of wind-waves: height $h$, steepness $\varepsilon$, age of waves A, and other characteristics of the wave spectrum (in dimensionless combinations of parameters), $U_d(h, \varepsilon, A...)$, could be expected on the basis of ratio (15).

According to the adopted model, all the sought dependencies are determined by the parametrizations for the drift-current gradient, $\partial U_d/\partial z$, and for the turbulent viscosity, $K_t$. The estimation of magnitude for the vertical gradient of drift velocity in the WZ, $\partial U_d/\partial z$, is completely based on the measurements, and it does not allow any functional changes. In this case, the explicit inverse dependence of $\partial U_d/\partial z$ on the wave height is compensated (in the adopted model) by the explicit linear dependence of the turbulent viscosity on $h$. Therefore, in this theory, the only degree of freedom, which allows the possibility for additional dependence of $K_t$ on the wave parameters, remains the specification of parametrization for dimensionless function $f_t(.)$. Thus, if the dependence of $f_t(.)$ on wave parameters is theoretically possible, then it appears a possibility for dependence of $U_d$ on the wave parameters, $U_d(h, \varepsilon, A, ...)$.

Since function $f_t(.)$ is responsible for the intensity of turbulent viscosity, it is necessary to look for those physical processes (and parameters) of wind waves which can change the degree of mixing in the WZ, and, therefore, affect the magnitude of turbulent viscosity coefficient in WZ, i.e. on the value of $f_t(.)$. The most likely processes of this kind can be: a) micro- and macro-breaking of wave crests (in terminology of paper by Longo et al, 2012a); and b) the shear instability of orbital wave motions. The intensity of the first process, unequivocally, must grow with the increase of wave steepness (Babanin, 2009); and the second does with the growth of wave amplitude (including the growing wave age) due to the increase of local Reynolds number for orbital wave motions. An increase of the breaking frequency is also possible while sharpening crests of waves, appearing with the age growth, due to the additional horizontal impact of wind on the crests presenting on the background of deep troughs (Babanin, 2009).

Now it should be noted that an increase of intensity for the vertical motions in the WZ corresponds, obviously, to a decrease in the effective viscosity, i.e. to the decrease of value for $f_t(.)$. Thus, on the basis of the said above, it can be assumed that function $f_t(.)$ will decrease with increasing both the steepness of waves, and, possibly, their age. As a result, according to formula (15), we can expect an increase of drift velocity $U_d$ with increasing steepness and wave age for a fixed value of the friction velocity.

It is impossible to predict theoretically the functional form of dependences $U_d(\varepsilon)$ and $U_d(A)$, because of the statistical nature of instability forming them. Besides, it should be remembered that with growing wave age $A$, both wave steepness $\varepsilon$ and depending on it friction velocity $u_{*a}$ tend to decrease (Drennan et al., 2003; Polnikov et al, 2003). Therefore, with increasing *A*, there are appearing the multidirectional trends, which can compensate each other to a large extent, canceling out potential dependence $U_d(A)$. Nevertheless, in the future, it seems quite reasonable to search for potentially possible direct empirical dependences of drift velocity both on steepness, $U_d(\varepsilon)$, and on wave age, $U_d(A)$, in addition to the explicit proportionality, Ud ~ $u_{*a}$, already established.

## 5. Conclusions

The results of the work allow us to draw the following conclusions.

5.1. In order to describe the drift velocity on a wavy water surface, the concept of the three-layer structure for the wind-wave-current system is introduced. The wave-zone (WZ) with dimensions of the order of two-three variance of wave record $\eta(t)$, in which the air and water are present alternately, has its own dynamics of processes in terms of average values. It is in this zone, the measured drift current is formed, which has a shear profile.

5.2. The experimentally established linear profile of the average drift current, $U(z)$, in the wave zone (Fig. 3) allows us to consider this zone as the analog of friction layer, in which a constant vertical flux of momentum is conserved, and constant coefficient of turbulent viscosity $K_t$ is realized.

5.3. The balance between the momentum flux and the vertical gradient of mean velocity, applied in the WZ, makes the basis of the drift current model (formula 11).

5.4. In the simplest case of independence of dimensionless function $f_t(.)$ (in formula (13) for turbulent viscosity $K_t$) of wave parameters, the model gives the linear proportionality between drift velocity on the water surface, $U_d$, and friction velocity in the air, $u_{*a}$, (formula 15).

5.5. In a more general case, formula (15) predicts a possible increase of drift velocity $U_d$ with an increase of mean wave slope $\varepsilon$ and their age *A*. This effect is physically stipulated by a decrease of turbulent viscosity (via function $f_t(.)$) with an increase of intensity of wave breaking, leading to increasing intensity of vertical movement dynamics.

5.6. The absence of direct empirical dependences of $U_d$ on wave parameters $\varepsilon$ and *A* (and others) requires its convincing empirical confirmation or their search on the basis of careful special experiments.

## 6. Acknowledgements

The project was supported by the RFBR grant No. 18-05-00161.